# Predicted superconductivity and superionic state in the electride Li$_5$N under high pressure

Zhongyu Wan[1], Chao Zhang[2], Tianyi Yang[1], Wenjun Xu[1], Ruiqin Zhang[1][3]*

[1] Department of Physics, City University of Hong Kong, Hong Kong SAR 999077, People's Republic of China

[2] Department of Physics, Yantai University, Yantai 264005, People's Republic of China

[3] Beijing Computational Science Research Center, Beijing 100193, People's Republic of China

*E-mail: aprqz@cityu.edu.hk

**Abstract**

Recently, electrides have received increasing attention due to their multifunctional properties as superconducting, catalytic, insulating, and electrode materials, with potential to offer other performance and possess novel physical states. This work uncovers that Li$_5$N as an electride possess four novel physical states simultaneously: electride state, super-coordinated state, superconducting state, and superionic state. By obtaining high-pressure phase diagrams of the Li-N system at 150-350 GPa using a crystal structure search algorithm, we find that Li$_5$N can remain stable as *P6/mmm* structure and has a 14-fold super-coordination number, as verified by Bader charge and electron localization function analysis. Its superconducting transition temperature reaches the highest at 150 GPa ($T_c$ = 48.97 K). Besides, Li$_5$N exhibits the superionic state at 3000 K, in which N atoms act like solid, while some Li atoms flow like liquid. The above results are further verified at a macroscopic level by using deep learning potential molecular dynamics simulations.

## 1. Introduction

As a special class of compounds[1,2], electrides involve quantum orbitals in the interstices of the lattice, and thus once the energy of their atomic orbitals is lower than that of the interstices, their electrons will leave the atom and enter the interstices[3]. Importantly, the presence of interstitial electrons primarily affects the physical and chemical properties of the crystals[2]. Since the anion electrons in the lattice are free from the constraints of the nucleus, the electride tends to provide electrons, leading to its application as a catalytic material with a relatively low work function[4], such as C12A7:(e$^-$)[5]. Some electrides have metallic characteristics and meet the requirements of electrode materials. For example, $Ca_2N$[6] and $Y_2C$[1] can be used as anode materials for Na cells, and their theoretical capacities can reach 1138 mAh/g[7] and 564 mAh/g[8]. In addition, electrides also exhibit novel physical properties due to the high concentration of electrons in the lattice making it superconducting at low temperatures, such as $Li_6P$[9], $Li_6C$[10], and $Li_7As$[11]. The electride NaLi[12] also can be used as an insulator material.

They are also divided into ambient-pressure electrides and high-pressure electrides depending on the environmental pressure. For electrides at ambient-pressure, such as C12A7:(e$^-$)[5], LaRuSi[13], LaScSi[14], LaCoSi[15], $Ba_2Ni_3$[16], and $Ca_3Pb$[17] were synthesized experimentally. But the high electronic activity makes it difficult to keep them stable at ambient-pressure[18]. This greatly limits the application of ambient-pressure electrides. Fortunately, high-pressure can make them stable[19], more and more electrides have been predicted in recent years[20], and they are often predicted to be superconducting materials[21].

Considering that high-pressure electrides are often consist of alkaline earth and non-metallic elements[22], this work investigates the binary Li-N system in the range of 150-350 GPa, aiming at discovering unknown novel states in electrides so as to provide theoretical guidance for experiments.

## 2. Computational Details

A crystal structure search for $Li_xN$ ($x$=1-9) at 150-350 GPa pressure using 1-3 times the molecular formula is performed using the particle swarm algorithm based CALYPSO software[23]. Structural optimization, band structure calculation, and molecular dynamics simulation are conducted using the Vienna Ab-initio Simulation Package (VASP)[24]. The projected-augmented wave pseudopotentials (PAW) is used to describe the electron-ion interaction[25], where $1s^22s^12p^0$ and $2s^22p^3$ are used as the valence electronic structures of Li and N, respectively. The generalized gradient approximation (GGA) and Perdew-Burke-Ernzerhof (PBE) are used as exchange-correlation generalizations[26]. The cutoff energy is set to 600 eV, k-point grid is divided at a spacing of $0.03\times2\pi$ using the Monkhorst-Pack method[27]. The relaxed ionic positions and cell parameters provided energy and forces smaller than $10^{-5}$ eV and 0.002 eV/Å, respectively. The ten lowest structures for each stoichiometric ratio are zero-point energy (ZPE) corrected due to quantum effects[28]. The phonon properties of a $3\times3\times3$ supercell are calculated based on density functional perturbation theory in combination with a PHONOPY code[29]. At the temperature T = 300, 1000, 2000, and 3000 K, a $5\times4\times3$ supercell with 360 atoms is used to perform ab initio molecular dynamics (AIMD) simulations for 8000 steps with a step of 1 fs under NVT condition. The QUANTUM ESPRESSO code is used to calculate the electron-phonon coupling (EPC) properties in the framework of linear response theory[30]. A ultra-soft standard solid-state pseudopotential with a cutoff energy of 100 Ry is used for Li and N. A dense k-point grid of $18\times18\times12$ with Methfessel-Paxton first-order spread set to 0.01 Ry, and first-order perturbation and kinetic matrix calculations are performed at irreducible points on a $3\times3\times2$ q-point grid centred on Γ point. The interatomic potential of the AIMD process with a cutoff radius of 8 Å and a three-layer network structure of [25, 50, 100] for 3 million iterations is trained by DEEPMD-KIT package[31], the samples are randomly divided into training set and test set in the ratio of 8:2. A $10\times10\times10$ supercell with 6,000 atoms performs 800,000 NPT simulations with a step of 1 fs. The temperature of 3000 K is controlled via the Nosé-Hoover thermostat, and the pressure of 150 GPa is maintained using the Parrinello-Rahman barostat as implemented in LAMMPS[32].

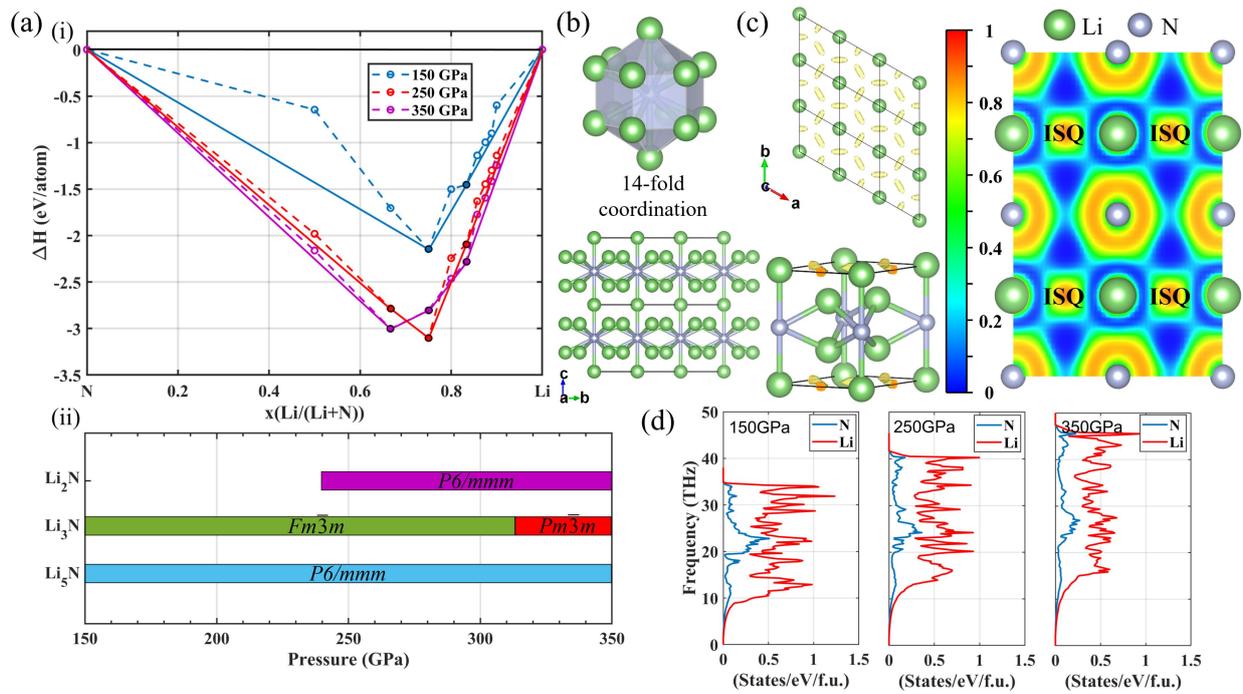

**Figure 1:** (a) Thermodynamic stability of the Li-N system at 150-350 GPa and phase diagrams of stable components. (b) Structural unit with 14-fold coordination and supercell structure. (c) Electron localization function of $Li_5N$ at 150 GPa (isosurface = 0.85). (d) Phonon projection density of states of $Li_5N$ at different pressures.

## 3. Results and Discussion

Although scientists have made structural predictions for N-Li compounds from 0-100 GPa, only $Li_3N_2$[33] has been found to behave as electride at 30 GPa, $Li_4N$[34] as a metastable electride at atmospheric pressure, and $LiN_5$[35] as a high energy density material. However, whether there are more novel states in other Li-N compounds at higher pressures remains unknown. In order to explore the possibility of Li-N forming stable compounds in the higher pressure interval, the enthalpy for each crystal is calculated to obtain the convex-hull diagram with ZPE correction at 150-350 GPa and the phase diagram of the stable components (Figure 1(a)):

$$\Delta H(Li_xN) = [H(Li_xN) - xH(Li) - H(N)]/(x+1) \quad (1)$$

where $H$ is the enthalpy of the most stable structure of certain compositions at the given pressure[36,37].

**Table 1. Structural information of the predicted stable Li-N phases from 150 to 350 GPa.**

| Compounds | Space group | Pressure (GPa) | Lattice Parameters (Å) | Atomic Wyckoff Positions (fractional) |
|---|---|---|---|---|
| $Li_5N$ | $P6/mmm$ | 150 | a=2.86<br>c=3.86 | N(1b) : (0, 0, 0.5)<br>Li(4h) : (0.33, 0.67, 0.29) |
| Li$_3$N | $Fm\bar{3}m$ | 150 | a=4.08 | N(4a) : (0, 0, 0) |
| | | | | Li(8c) : (0.25, 0.25, 0.25) |
| Li$_2$N | $P6/mmm$ | 250 | a=2.66 | Li(4b) : (0.5, 0, 0) |
| | | | | N(1b) : (0, 0, 0.5) |
| | | | c=1.79 | Li(2c) : (0.67, 0.33, 0) |
| Li$_3$N | $Pm\bar{3}m$ | 350 | a=2.38 | N(1b) : (0.5, 0.5, 0.5) |
| | | | | Li(3d) : (0.5, 0, 0) |

Li$_3$N and Li$_5$N are always located on the convex-hull line, indicating that they do not decompose into monomers and other compounds thermodynamically. Li$_5$N has only 1 stable high-pressure phase, *P6/mmm*, with each N atom surrounded by 14 Li atoms to form an 18-fold structural unit. The whole crystal is formed in layers perpendicular to the c-axis (Figure 1(b)), similar to graphene. Li$_3$N has two high-pressure phases in the interval, and as the pressure increases, Li$_3$N transforms from face-centered cubic ($Fm\bar{3}m$) to body-centered cubic ($Pm\bar{3}m$) at 315 GPa. In addition, Li$_2$N is stable at 240 GPa in a layered structure with *P6/mmm*. Furthermore, experimental α-Li$_3$N at atmospheric pressure[38] is also confirmed (*P6/mmm*) in this computational study, which validates the crystal structure search. Since the *P6/mmm* Li$_5$N has ultra-high 14-fold coordination, there may be novel states in this compound, and the Bader charge[39] is used to analyze the valence of the atoms in the lattice (N = +2.31e, Li = -0.47e, 150 GPa). It is worth noting that Bader charges can only qualitatively determine the atomic charge, as they have a slightly smaller value than the nominal ionic charge[40] in which each N atom gains 3 electrons and the Li atom loses 1 electron. The electron localization function (ELF)[41] demonstrates the destination of the surplus electrons (Figure 1(c)), which are clustered in the interstices of the 2D plane formed by the lithium atoms on (001) and surrounded by two Li atomic layers. Because the N atom has gained three electrons and reached full 2p orbitals, it cannot accommodate the excess Li 2s electrons, and the high pressure makes the distance between the atoms shorter. The average distance of the neighboring Li-Li around the interstitial electrons is 1.89 Å, which means that there is a strong Coulomb repulsion between the Li atoms considering that the Li atoms have a radius of 1.09 Å. To make the lattice stable, the 2s electrons leave the Li atoms and enter the interstices to form an interstitial quasi-atom (ISQ)[42].

In addition to the thermodynamic stability, the stability of lattice dynamics of the crystal is also necessary to be satisfied. The phonon density of states (PHDOS) of $Li_5N$ at 150-350 GPa reveals (Figure 1(d)) that there are no electronic states at imaginary frequencies, which indicates that $Li_5N$ is dynamically stable. In addition, the Li and N atoms present similar trends in PHDOS, showing a strong coupling effect between them. According to Liu's work[10], this implies that $Li_5N$ could be a potential superconductor. To verify whereas the superconductivity depends on its electronic structure, insight into the contribution of ISQ to the electronic structure is obtained by inserting pseudo-atoms at the fractional coordinates (0.5, 0, 0), (0, 0.5, 0) and (0.5, 0.5, 0).

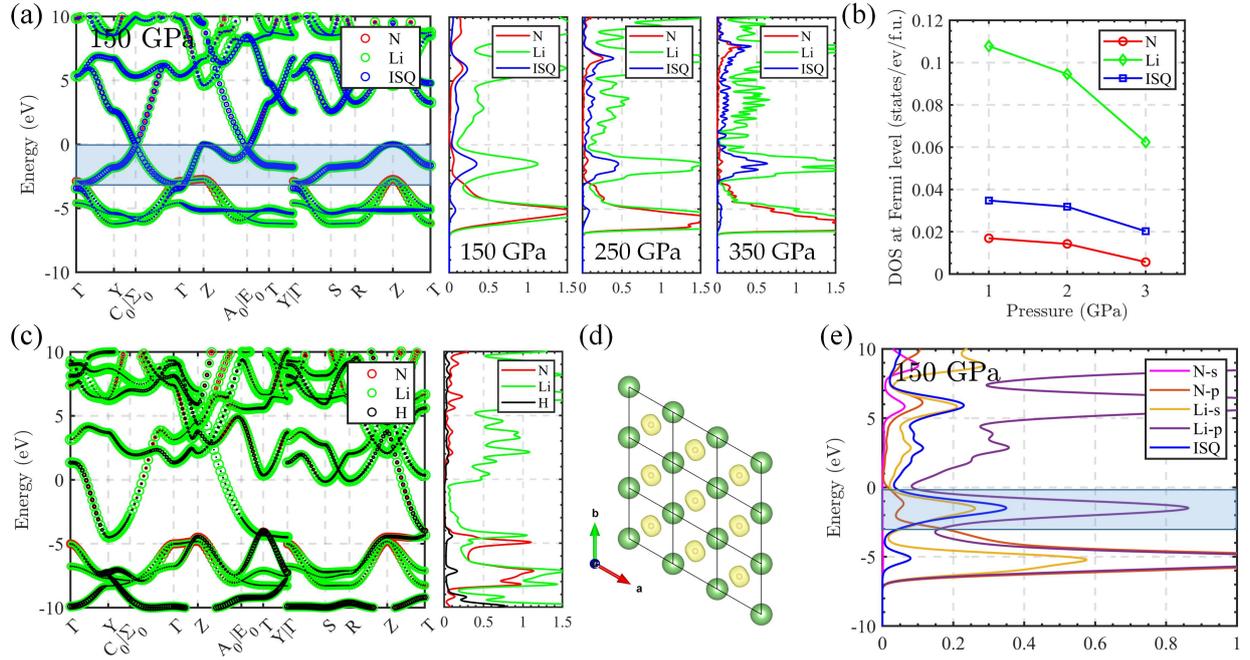

**Figure 2. (a) Projected band structure of $Li_5N$ at 150 GPa, and density of states at different pressures. (b) Density of states of individual atoms on the Fermi-level. (c) Projected band structure of $Li_5N$ at 150 GPa after insertion of a H atom. The size of the circles represents the weights of different atoms. (d) Electron localization function of $Li_5NH$ (isosurface = 0.85). (e) Density of states with angular momentum decomposition at 150 GPa.**

Figure 2(a) shows the projected band structure of $Li_5N$ at 150 GPa. It can be seen that the band near the Fermi level is mainly composed of Li and ISQ. Although the green bubble representing the Li atom is the largest, the blue bubble representing the ISQ also has a significant weight. This suggests an important contribution of interstitial electrons to the metallization of the electride $Li_5N$, which is very similar to the behavior of interstitial electrons in $Y_2C$ and $Ca_2N$[6]. The overlap of ISQ and Li in the band near the Fermi level also proves that the interstitial electrons originate from Li atoms. Furthermore, the electron density of states (DOS) reveals the impact of pressure effects on the electronic distribution, where N and Li have great density of states and similar trends at the energy level (-4,-6), which implies the overlap of wave functions between N and Li, it leads to the formation of stable N-Li bonds, thus occupying a lower energy level. Notice that the electronic state of Li has a distinct DOS peak at the higher energy level of -2 eV, but the DOS of N is not obviously displayed. The excess Li atoms in the system may be unable to form chemical bonds with N atoms, making it impossible for the excess Li to occupy lower energy levels. In addition, the superconductivity of a compound is usually correlated with the electronic states at the Fermi level. Figure 2(b) demonstrates the effect of pressure on the distribution of electrons at the Fermi level, where an increase in pressure results in less occupation of electronic states in N, Li, and ISQ, which means that fewer electrons can form Cooper pairs, leading to a weakening of superconductivity.

Since interstitial electrons have anionic nature, they are easily attracted to atoms with strong electronegativity, which means that the insertion of cations will attract interstitial electrons, making the system no longer an electride and the electron distribution in the lattice is greatly altered, significantly affecting its electronic structure[43]. To obtain the origin of the ISQ, H atoms are inserted in fractional coordinates (0.5, 0.5, 0) to compare the electronic structures of the conventional compound $Li_5NH$ and the electride $Li_5N$. Figure 2(c) shows the electronic structure of $Li_5N$ at 150 GPa after being inserted into the H atom. Compared with Figure 2(a), the band of the anion near the Fermi level disappears because the original interstitial electrons are attracted by H.

On the contrary, more 1s orbitals of H appear at the energy level (-10,-5). This result indicates the transfer of interstitial electrons to the H atom, which further proves the existence of ISQ. In contrast to Figure 1(c), all the interstitial electrons scattered in the lattice are transferred to the H atom, which no longer possesses superconductivity because of the Jahn-Teller effect[44] induced by the concentrated charge of H atom, causing $Li_5NH$ to transform into a kinetically unstable conventional compound. In the electride $Li_5N$, its superconductivity may be electride induced because the contribution of ISQ near the Fermi level allows it to be superconducting.

Since the interstitial electron can be viewed as a quasi-atom, it must interact with other atoms in the system. ISQ mainly occupies the energy band in the interval (0,-3) (Figure 2(a)), and Figure 2(e) shows the DOS with angular momentum decomposition. It can be seen that the DOS of N-p, Li-s, Li-p, and ISQ have great values and the same trend in this energy interval, which implies the existence of sp orbital hybridization between them.

The calculated EPC properties at different pressures reveal that the largest EPC coefficient ($\lambda$=1.39) at 150 GPa is very close to the superconductor $Li_3S$ ($\lambda$=1.43)[45]. The superconducting critical temperature ($T_c$) can be derived via the McMillan-Allen-Dynes equation:

$$T_c = \frac{\omega_{\log}}{1.2} \exp\left[-\frac{1.04(1+\lambda)}{\lambda - \mu^*(1+0.62\lambda)}\right] \quad (2)$$

where $\omega_{\log}$ is the log-averaged phonon frequency, $\mu^*$ is the Coulomb pseudopotential ($\mu^*$ = 0.1) and $\lambda$ is the EPC constant. Its $T_c$ (150 GPa, $T_c$=48.97 K) exceeds $Li_6P$[9] (270 GPa, $T_c$=39.3 K), which is also an electride. The superconducting critical temperature of $Li_5N$ decreases with increasing pressure, in contrast to the response of the other electride superconductor $Li_6C$[10]. According to the phonon dispersion curve[46] and the Eliashberg spectral function[47] shown in Figure 3(a), phonons below 10 THz at 150 GPa contribute 58.5% to the EPC constant, indicating that the low-frequency phonon modes (especially at the Z point) determine the $T_c$ of $Li_5N$. In addition, the PHDOS (Figure 1(d)) shows a tendency for phonons to migrate from low to high frequencies, resulting in fewer phonons in the low-frequency state, which may account for the weakened $\lambda$.

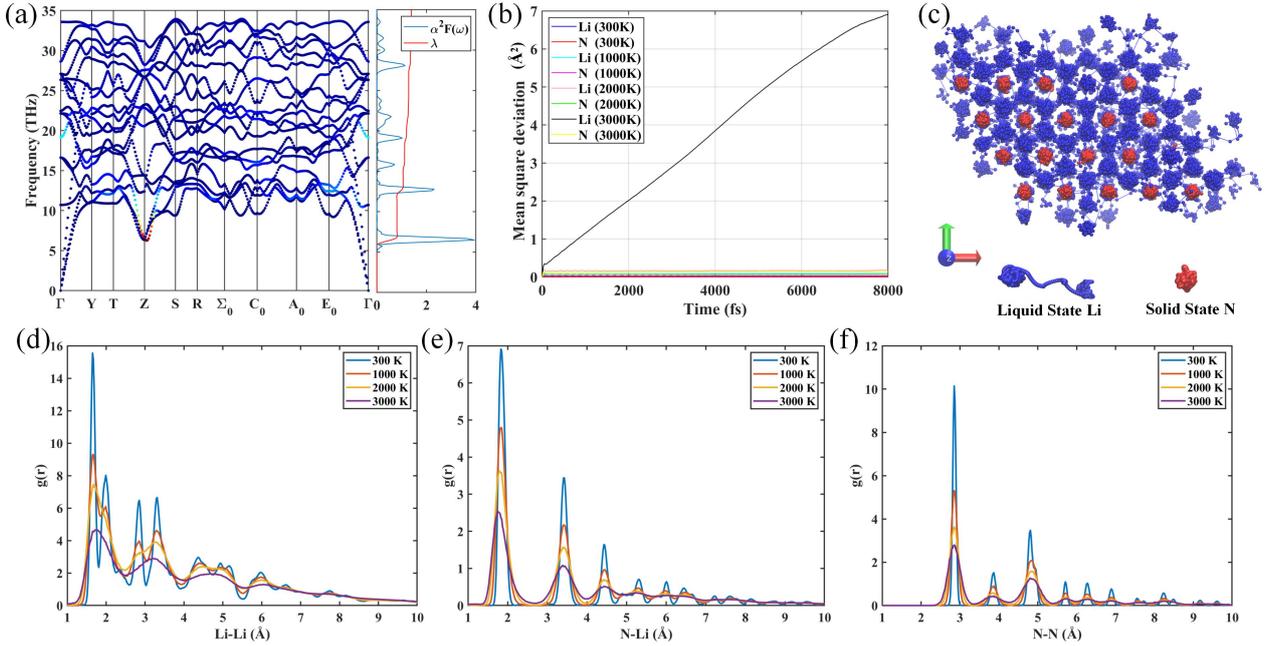

**Figure. 3.** (a) The phonon dispersion curve and Eliashberg spectral function at 150 GPa, and the brightness of each point color in the phonon dispersion curve is proportional to the EPC weight ($\lambda_{q,v}$). (b) The mean square displacement curve of atoms during AIMD at different temperatures. (c) Trajectory at 3000 K (Li in blue, N in red). (d) RDF curves of Li-Li. (e) RDF curves of N-Li. (f) RDF curves of N-N.

**Table 2. Superconductivity of $Li_5N$ at different pressures**

| Pressure (GPa) | $\lambda$ | $\omega_{\log}$ (K) | $T_c$ (K) |
| --- | --- | --- | --- |
| 150 | 1.39 | 463.42 | 48.97 |
| 250 | 0.37 | 748.19 | 1.95 |
| 350 | 0.19 | 767.19 | 0.01 |

Since the DFT calculations are based on 0 K temperature, which is far from the actual environment, it is necessary to consider the effect of different temperatures on the $Li_5N$ structure. The AIMD results (Figure 3(b)) demonstrate that the structure of $Li_5N$ is stable at 300, 1000, and 2000 K. The slope of the mean square displacements (MSD) curves ($k≈0$) prove the presence of the atoms in the system as solid state. However, the high temperature of 3000K allow $Li_5N$ to exhibit novel physical states[48,49]. Surprisingly, the trajectories of all the atoms (Figure 3(c)) show that the N atoms and some of the Li atoms in them vibrate near their initial positions, exhibiting the properties of solids. Interestingly, a few Li atoms flow in chains and behave more like liquids, suggesting the presence of a superionic state[50] in $Li_5N$. The observation is verified by the calculated MSD, where the MSD of N remains almost constant ($k_N ≈ 0$), while the MSD of Li continues to increase ($k_{Li} ≈ 0$). Compared to other compounds that exhibit superionic states at about 1000 K[48,49], $Li_5N$ requires higher temperatures because Li has a larger atomic mass than He and H. The energy homogeneity theorem proves that energy at equilibrium is evenly distributed between the degrees of freedom and the displacement of the atoms is stronger due to the lighter mass under the same kinetic energy conditions. The results of the radial distribution function (RDF) verify the superionic behavior at 3000 K (Figure 3 (d)-(f)). Whether it is Li-Li, N-Li or N-N, it can be seen that the height of the peak decreases and becomes wider as the temperature increases, indicating that higher temperatures make the system change from solid to amorphous state.

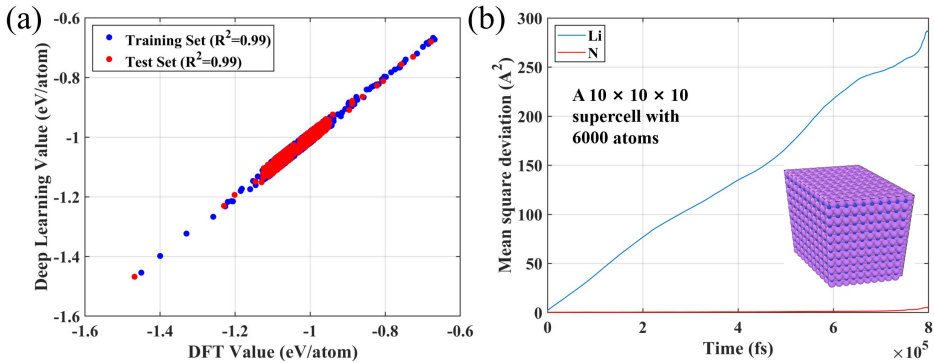

**Figure 4:** (a) Scatter plot of the energy of DFT versus deep learning; (b) MSD of the deep potential molecular dynamic (DPMD) process.

It is worth noting that the simulated system contains only 360 atoms. And compared to NVT, the NPT simulations are much closer to the actual conditions, and the simulation time is too short. Based on this consideration, we use deep learning to train new interatomic potential functions to scale up the AIMD system (360 atoms, 8 ps) to an actual condition (6000 atoms, 800 ps). It can be seen that the coefficient of determination ($R^2$) of the test set is very close to 1, demonstrating that the deep network exhibits strong generalization capabilities (Figure 4(a)). The $R^2$ of the training set and test set are also very close, which indicates that the problem of over-fitting is overcome during the training process. The DPMD result shows a similar trend to AIMD, suggesting that the superionic state in $Li_5N$ can be present at the macroscopic scale.

## 4. Conclusion

In this work, the potential stable structures of the Li-N system at high pressures of 150-350 GPa are investigated using crystal structure search algorithms combined with first-principles calculations, and the high-pressure phase diagrams of $Li_2N$, $Li_3N$, and $Li_5N$ are determined. $Li_5N$ can persist as a *P6/mmm* structure and forms a 14-fold coordinated electride, which exhibits metallicity and superconductivity at 150 GPa. It has the largest $T_c$ among all known electrides (48.97 K). Its $T_c$ decreases with increasing pressure, and ISQ plays a significant role in maintaining the metallicity of $Li_5N$. In addition, the compound also exhibits a superionic state at temperature of 3000K, where the N immobile while some Li atoms flow like liquid. The superionic states at the macroscopic scale are further verified by performing DPMD simulations.

**Acknowledgement**

The work described in this paper was supported by grants from the Research Grants Council of the Hong Kong SAR (CityU 11305618 and 11306219), the National Natural Science Foundation of China (11874081 and 11874318）.